\documentclass[]{aa} 
\usepackage{epsfig, graphicx}

\begin{document}

%\thesaurus{06(08.01.1; 08.03.02; 08.16.2)}

%\titlerunning{} 

\title{Beryllium abundances in stars hosting giant planets\thanks{Based on
    observations collected with the VLT/UT2 Kueyen telescope (Paranal Observatory, 
    ESO, Chile) using the UVES spectrograph (Observing runs 66.C-0116\,A and 
    66.D-0284\,A), and with the William Herschel and Nordic Optical Telescopes, 
    operated at the island of La Palma by the Isaac Newton Group and jointly by Denmark, 
    Finland, Iceland, and Norway, respectively, in the Spanish Observatorio del Roque de 
    los Muchachos of the Instituto de Astrof\'{\i}sica de Canarias.}}

%\subtitle{}

\author{N.C.~Santos\inst{1} \and R.J.~Garc\'{\i}a L\'opez\inst{2,3} \and 
G.~Israelian\inst{2} \and M.~Mayor\inst{1} \and R.~Rebolo\inst{2,4} \and 
A.~Garc\'{\i}a-Gil\inst{2} \and M.R.~P\'erez de Taoro\inst{2} 
\and S.~Randich\inst{5} } 

\offprints{Nuno C. Santos, \email{Nuno.Santos@obs.unige.ch}}

\institute{Observatoire de Gen\`eve, 51 ch.  des
	   Maillettes, CH--1290 Sauverny, Switzerland \and 
	   Instituto de Astrof{\'\i}sica de Canarias, E-38200 
	   La Laguna, Tenerife, Spain \and 
	   Departamento de Astrof\'{\i}sica, Universidad de La Laguna, 
	   Av. Astrof\'{\i}sico Francisco S\'anchez s/n, E-38206, La 
	   Laguna, Tenerife, Spain \and 
	   Consejo Superior de Investigaciones Cient\'{\i}ficas, Spain \and
	   INAF/Osservatorio Astrofisico di Arcetri, Largo Fermi 5, I-50125 
	   Firenze, Italy}

\date{Received  /  Accepted } 

\titlerunning{Beryllium abundances in stars hosting giant planets} 

%--------------------------------------------------------------------------

\abstract{
     We have derived beryllium
     abundances in a wide sample of stars hosting planets, with
     spectral types in the range F7V$-$K0V, aimed at studying in
     detail the effects of the presence of planets on the structure and
     evolution of the associated stars. Predictions from current
     models are compared with the derived abundances and
     suggestions are provided to explain the observed 
     inconsistencies. We show that while still not
     clear, the results suggest that theoretical models
     may have to be revised for stars with T$_\mathrm{eff}<$5500\,K.
     On the other hand, a comparison between planet host and non-planet host stars 
     shows no clear difference between both populations. 
     Although preliminary, this result favors a ``primordial'' origin for 
     the metallicity ``excess'' observed for the planetary host stars.
     { Under this assumption, i.e. that there would be no differences between
     stars with and without giant planets, the light element depletion pattern of our 
     sample of stars may also be used to further investigate and constraint 
     Li and Be depletion mechanisms.}
\keywords{stars: abundances -- 
          stars: chemically peculiar -- 
          stars: evolution --
	  planetary systems
	  }
}

\maketitle

\section{Introduction}
\label{sec:introduction}

In the last few years we have witnessed a fantastic development in an ``old''
but until recently not very successful field of astrophysics: the search for
extra-solar planets. Following the first success in exoplanet searches 
with the discovery of the planet around \object{51\,Peg} 
(Mayor \& Queloz \cite{May95}), the number of known giant planets orbiting solar-type stars
did not stop growing, being currently of 72 (including 7 multi-planetary 
systems)\footnote{Counts of January 2002. For a list see e.g. table at
{\tt obswww.unige.ch/$\sim$naef/who\_discovered\_that\_planet.html}.}. 

Unexpectedly, the planets found to date do not have much in common 
with the ones in our own Solar System (for a review see
Marcy et al. \cite{Mar00}, Udry et al. \cite{Udr01}, or 
Mayor \& Santos \cite{May01b}). One remarkable characteristic appears to
be related with the parent stars themselves: stars with planetary companions 
are considerably metal-rich when compared with single field
dwarfs (Gonzalez \cite{Gon98}; Santos et al. \cite{San00}; Gonzalez et 
al. \cite{Gon01}; Santos et al. \cite{San01}, \cite{San01b}). To explain the
observed difference two main explanations have been suggested. The 
first and more ``traditional'' is based upon the fact that the more metals 
you have in the proto-planetary disk, the higher should be the probability of forming a
planet (see e.g. Pollack et al. \cite{Pol96} for the traditional paradigm of 
planetary formation). Thus, in this case the ``excess'' of metallicity is 
seen as primordial to the cloud that gave origin to the star/planet system. 
``Opposing'' to this view, the high metal 
content observed for stars with planets has also been interpreted as a sign of the 
accretion of high-Z material by the star sometime after it reached the 
main-sequence (e.g. Gonzalez \cite{Gon98}; Laughlin \cite{Lau00}).

\begin{figure}[t]
\psfig{width=\hsize,file=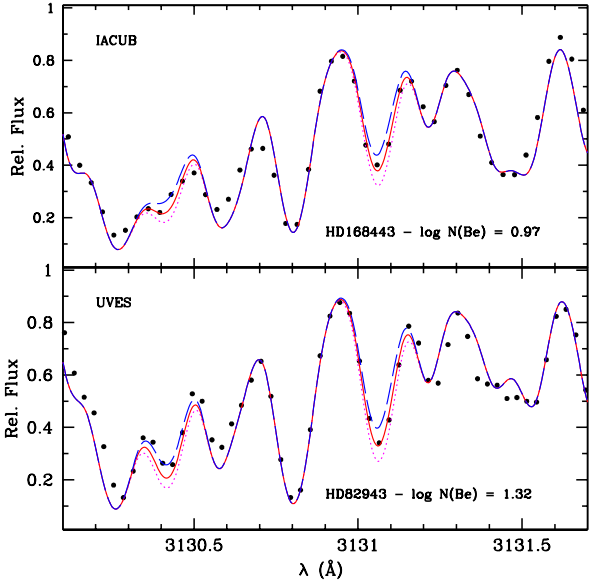}
\caption[]{Spectra in the \ion{Be}{ii} line region (dots) for two of the objects observed, 
and three spectral synthesis with different Be abundances, corresponding to the
best fit (solid line) and to changes of $\pm$0.2\,dex, respectively.}
\label{fig1}
\end{figure}

\begin{figure}[t]
\psfig{width=\hsize,file=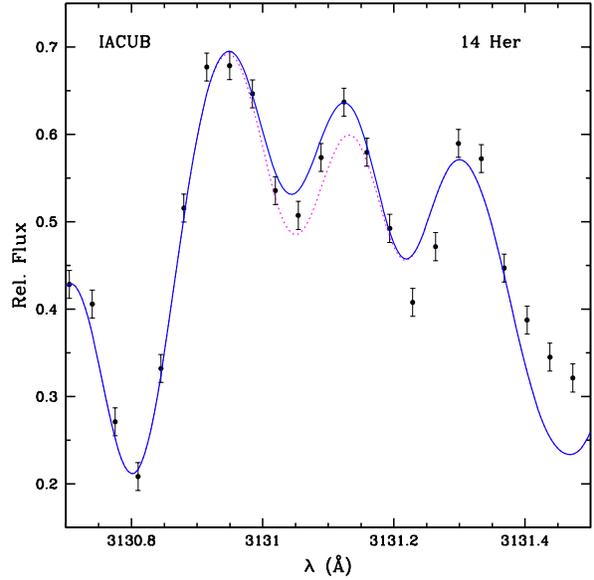}
\caption[]{Spectrum in the \ion{Be}{ii} line region (dots) for \object{14\,Her} 
(\object{HD\,145675}), and two spectral synthesis with different Be abundances, corresponding to no Be
(solid line) and to $\log{N(Be)}$=0.5. Error-bars represent the photon noise error.
A conservative upper limit of 0.5 for the Be abundance was considered.}
\label{fig2}
\end{figure}

Although recent results seem to favor the former scenario as the key process
leading to the observed metal richness of stars with planets 
(Santos et al. \cite{San01}, \cite{San01b}; Pinsonneault et al. \cite{Pin01}), signs of accretion of 
planetary material have also been found for some planet hosts (e.g. Israelian et al. \cite{Isr01a}; 
Laws \& Gonzalez \cite{Law01}). The question is then turned to know how frequent those phenomena happen, 
and to how much these could have affected the observed metal contents.

{ One possible and interesting approach to this problem may pass by the
study of one particularly important element: beryllium (Be). Together with lithium 
(Li) and boron (B), Be is a very important tracer of the internal stellar structure and 
kinematics. Be is mainly produced by spallation reactions in the
interstellar medium, while it is burned in the hot stellar furnaces 
(e.g. Reeves \cite{Ree94}). While most works of light element abundances are based on Li (the abundances 
of this element are easier to measure),} Be studies have one major advantage when 
compared with Li. Since it is burned at much higher temperatures, 
Be is depleted at lower rates than Li, and thus we can expect to measure Be 
abundances in stars which have no detectable Li in their atmospheres
(like intermediate-age late G- or K-type dwarfs). 
In fact, for about 50\% of the known planet host stars no Li was detected 
(Israelian et al. \cite{Isr01b}). { Furthermore, Li studies have shown
the presence of a significant scatter for late-type stars of similar temperature. This has also
been observed in open clusters where all stars have the same age, and appears to be related
to the clusters's age, rotational velocities, pre-Main Sequence history, etc. (see, 
e.g., Garc\'{\i}a L\'opez et al. \cite{Gar94}; Randich et al. \cite{Ran98}; Jones et al. \cite{Jon99}), 
a fact that may complicate or even preclude a comparison.}

Given all these points, Be studies of planetary host stars
can indeed be particularly important and telling. 
{ For example,} if pollution has played some important role in determining the 
high-metal content of planet host stars, we would expect to find a similar or even 
higher increase in the Be contents. { This is basically due to the fact that
planetary material is relatively poor in H and He when compared
to the star (e.g. Anders \& Grevesse \cite{And89}). Furthermore, and unlike for iron, Be 
may be already a bit depleted
in the stellar surface. Thus, the injection of planetary material into this latter
could even be responsible for a more important abundance change in the Be abundance than in the
iron content. If pollution has indeed played an
important role, the net result of
the fall of planetary material into the central ``sun'' would thus be that, for a given 
temperature interval, we 
should find that planet hosts are (in principle) more Be-rich that non-planet host 
stars. In other words, the analysis of Be abundances represents an independent way of testing 
the pollution scenario.}

Garc\'{\i}a L\'opez \& P\'eres de Taoro (\cite{Gar98}) carried out the first
Be measurements in stars hosting planets: \object{16\,Cyg\,A} and B, and \object{55\,Cnc},
followed by Deliyannis et al. (\cite{Del00}).
In order to continue to address this problem, we present\footnote{Preliminary results were already
presented in Garc\'{\i}a L\'opez et al. (\cite{Gar01}).} here a study of 
Be abundances in a set of 29 stars with planets, and a smaller set of 6 stars without known
planetary companions. In Sects.\,\ref{sec:observations} and \ref{sec:analysis} 
we present the observations and analysis of the data and in Sect.\,\ref{sec:be} we discuss the results 
{in the context of the planetary host stars chemical abundances, but also in terms
of the Be depletion processes.} We conclude in Sect.\,\ref{sec:conclusions}. 

\section{Observations and data reduction}
\label{sec:observations}

Observations of the \ion{Be}{ii} doublet at 3131\,\AA\ were carried out during 
5 different observing runs. Two of them made use of the UVES spectrograph, at the VLT/UT2 Kueyen 
telescope (ESO, Chile). The obtained spectra have a resolution 
R$\equiv\lambda/\Delta\lambda$ of around 70\,000 for one of the runs 
(66.C-0116\,A) and 50\,000 for the other (66.D-0284\,A), and the S/N 
ratios in the region around 3131\,\AA\ varying between 30 and 250 (from now on we will 
refer to runs UVES(A)/(B), respectively). UVES 
spectra were complemented with data obtained using the UES 
spectrograph (R=55\,000) at the 4.2-m William Herschel Telescope 
(WHT) and, in two different observing runs (A/B), with the IACUB echelle spectrograph 
(R=33\,000) at the 2.6-m Nordic Optical Telescope (NOT), 
both at the Observatorio del Roque de los Muchachos (La Palma). The observations, together 
with the corresponding observing run and S/N ratios obtained are listed in Table~\ref{tab1}.

Data reduction was done using IRAF\footnote{IRAF 
is distributed by National Optical Astronomy Observatories, operated 
by the Association of Universities for Research in Astronomy, Inc., 
under contract with the National Science Foundation, U.S.A.} tools 
in the {\tt echelle} package. Standard background correction, 
flat-field, and extraction procedures were used. For the UVES(A) 
run and for the UES data, the wavelength calibration was 
done using a ThAr lamp spectrum taken during the same night. For 
the UVES(B) and IACUB runs the wavelength calibration was 
done using photospheric lines in the region of interest. 
The final linear dispersions for the UVES(A), UVES(B), IACUB(A), IACUB(B), 
and UES spectra were 30, 17, 35, 35 and 17 m\AA\,pixel$^{-1}$, respectively, 
with rms values in the range 4$\times10^{-4}$ to 8$\times10^{-3}$\,\AA.

\section{Abundance Analysis} 
\label{sec:analysis}

\subsection{Atmospheric parameters}

Garc\'{\i}a L\'opez et al. (\cite{Gar95}) carried out a detailed analysis of
the sources of uncertainties regarding the determination of Be abundances. They 
concluded that the precision of the
derived Be abundances is mostly dependent on the choice of the 
stellar atmospheric parameters. In particular, they are very sensitive 
to the adopted value for the surface gravity ($\log{g}$). 

In order to limit the possible systematic errors in our 
determinations it is important, whenever possible, to use an 
uniform set of atmospheric parameters for all 
the programme stars. We thus decided to use the values listed by Santos 
et al. (\cite{San01}; \cite{San01b}), computed from an uniform and accurate 
spectroscopic analysis available for most of the stars studied in this 
paper\footnote{The list of atmospheric parameters in Santos et al. (\cite{San01}) was determined by 
these authors for part of the objects, or was taken from the studies of Gonzalez (\cite{Gon98}), 
Gonzalez \& Laws (\cite{Gon98}), and Gonzalez et al. (\cite{Gon99},\cite{Gon01}).}.
For three of the stars (\object{HD\,870}, \object{HD\,1461}, and \object{HD\,3823} for which 
no planetary companions were found to date), no 
parameters were available, and we computed them using CORALIE or FEROS spectra, 
in the very same way as in Santos et al. (\cite{San01}). The values are listed in 
Table~\ref{tab2}. As discussed by the authors, the errors in 
$T_{\mathrm{eff}}$ are usually lower than 50\,K, and errors in the microturbulence 
parameter are of the order of 0.1\,km\,s$^{-1}$. As for $\log{g}$, the uncertainties are in 
the range of 0.10 to 0.15\,dex\footnote{Note that these small uncertainties are relative, 
and not absolute.}.

\begin{table*}[ht]
\caption[]{Observing log for stars with and without planets.}
\begin{tabular}{llrlrr}
\hline
\multicolumn{1}{l}{HD} & \multicolumn{1}{l}{Star} & \multicolumn{1}{l}{V} & \multicolumn{1}{l}{Observ.} & \multicolumn{1}{l}{S/N} & \multicolumn{1}{l}{Date}\\
\multicolumn{1}{l}{number} & \multicolumn{1}{l}{} & \multicolumn{1}{l}{} & \multicolumn{1}{l}{Run} & \multicolumn{1}{l}{}
& \multicolumn{1}{l}{}\\
\hline\\[-0.2cm]
\multicolumn{4}{l}{Stars with planets:}\\[0.1cm]
\object{HD\,13445}  &Gl\,86         & 6.1 &UVES(A)  &150   &  Nov. 2000\\
\object{HD\,16141}  &HD\,16141      & 6.8 &UVES(A)  &120   &  Nov. 2000\\
\object{HD\,17051}  &$\iota$\,Hor   & 5.4 &UVES(A)  &150   &  Nov. 2000\\
\object{HD\,52265}  &HD\,52265      & 6.3 &UVES(A)  &120  &  Dec. 2000\\
\object{HD\,75289}  &HD\,75289      & 6.4 &UVES(A)  &110  &  Dec. 2000\\
\object{HD\,82943}  &HD\,82943      & 6.5 &UVES(A)  &140  &  Jan. 2001\\
\object{HD\,210277} &HD\,210277     & 6.5 &UVES(A)  &110   &  Nov. 2000\\
\object{HD\,217107} &HD\,217107     & 6.1 &UVES(A)  &120   &  Nov. 2000\\
\multicolumn{1}{c}{--} &BD$-$10\,3166 &10.0 &UVES(B)  &20    &  Feb. 2001\\
\object{HD\,38529 } &HD\,38529      & 5.9 &UVES(B)  &60    &  Feb. 2001\\
\object{HD\,75289}  &HD\,75289      & 6.4 &UVES(B)  &30  &  Feb. 2001\\
\object{HD\,92788}  &HD\,92788      & 7.3 &UVES(B)  &40    &  Feb. 2001\\
\object{HD\,82943}  &HD\,82943      & 6.5 &UVES(B)  &35  &  Feb. 2001\\
\object{HD\,108147} &HD\,108147     & 7.0 &UVES(B)  &60    &  Feb. 2001\\
\object{HD\,121504} &HD\,121504     & 7.5 &UVES(B)  &45    &  Feb. 2001\\
\object{HD\,134987} &HD\,134987     & 6.5 &UVES(B)  &60    &  Feb. 2001\\
\object{HD\,95128}  &47\,UMa        & 5.1 &IACUB(A)    &100   &  May 2000\\
\object{HD\,114762} &HD\,114762     & 7.3 &IACUB(A)   &65    &  May 2000\\
\object{HD\,117176} &70\,Vir        & 5.0 &IACUB(A)    &70    &  May 2000\\
\object{HD\,130322} &HD\,130322     & 8.0 &IACUB(A)    &35    &  May 2000\\
\object{HD\,145675} &14\,Her        & 6.7 &IACUB(A)    &65    &  May 2000\\
\object{HD\,168443} &HD\,168443     & 6.9 &IACUB(A)    &55    &  May 2000\\
\object{HD\,187123} &HD\,187123     & 7.9 &IACUB(A)    &55    &  May 2000\\
\object{HD\,195019} &HD\,195019     & 6.9 &IACUB(A)    &50  &  May 2000\\   
\object{HD\,10697}  &109\,Psc      & 6.3 & IACUB(B)   &40    & Oct. 2001\\
\object{HD\,12661}  &HD\,12661      & 7.4 & IACUB(B)   &40    & Oct. 2001\\
\object{HD\,22049}  &$\epsilon$ Eri & 3.7 & IACUB(B)   &100    & Oct. 2001\\
\object{HD\,9826}  &$\upsilon$\,And & 4.1 &UES      &120   &  Aug. 1998\\
\object{HD\,120136} &$\tau$\,Boo    & 4.5 &UES      &90    &  Aug. 1998\\
\object{HD\,143761} &$\rho$\,CrB    & 5.4 &UES      &120   &  Aug. 1998\\
\object{HD\,217014} &51\,Peg        & 5.5 &UES      &100   &  Aug. 1998\\[0.15cm]
\hline\\[-0.2cm]
\multicolumn{4}{l}{Stars without known planets:}\\[0.1cm]
\object{HD\,870}    & HD\,870     & 7.2 &UVES(A)    &130    &  Nov. 2000\\
\object{HD\,1461}   & HD\,1461    & 6.5 &UVES(A)    &120    &  Nov. 2000\\
\object{HD\,1581}   & HD\,1581    & 4.2 &UVES(A)    &140   &  Dec. 2000\\
\object{HD\,3823}   & HD\,3823    & 5.9 &UVES(A)    &130    &  Oct. 2000\\
\object{HD\,26965A}& $o^2$\,Eri     & 4.4 &IACUB(B) &55    &  Oct. 2001\\
\object{HD\,222335} & HD\,222335  & 7.2 &UVES(A)    &110   &  Dec. 2000\\
\hline
\end{tabular}
\label{tab1}
\end{table*}

\begin{table*}[t]
\caption[]{Stellar atmospheric parameters and resulting beryllium abundances for each spectrum.}
\begin{tabular}{lllrclc}
\hline
\multicolumn{1}{l}{Star} & \multicolumn{1}{l}{T$_\mathrm{eff}$} & \multicolumn{1}{l}{ $\log{g}$} & \multicolumn{1}{r}{[Fe/H]} & \multicolumn{1}{c}{$\log{N(Be)}$} & \multicolumn{1}{l}{Run} & \multicolumn{1}{r}{$\log{N(Li)}\dag$} \\
\hline\\[-0.15cm]
\multicolumn{7}{l}{Stars with planets$\dag\dag$:}\\[0.1cm]
\object{BD\,$-$10\,3166} &6320 &4.38 &0.33 &$<$0.55 &UVES(B) &\multicolumn{1}{c}{--} \\
\object{HD\,9826} &6140 &4.12 &0.12 &0.99$\pm$0.18 &UES &2.26 \\
\object{HD\,10697} & 5605 & 3.96 & 0.16 &1.38$\pm$0.18 & IACUB(B) & 1.94\\
\object{HD\,12661} & 5715 & 4.45 & 0.35 &1.07$\pm$0.14 & IACUB(B) & $<$0.99\\
\object{HD\,13445} &5205 &4.70 &$-$0.20 &$<$0.52 &UVES(A) &$<$0.5 \\
\object{HD\,16141} &5805 &4.28 &0.15 &1.27$\pm$0.14 &UVES(A) &$<$0.73 \\
\object{HD\,17051} &6225 &4.65 &0.25 &1.02$\pm$0.17 &UVES(A) &2.63 \\
\object{HD\,22049} & 5135 &4.70 &$-$0.07&0.70$\pm$0.20 & IACUB(B) & $<$0.3\\
\object{HD\,38529} &5675 &4.01 &0.39 &$<$0.30 &UVES(B) &$<$0.61 \\
\object{HD\,52265} &6100 &4.29 &0.24 &1.21$\pm$0.14 &UVES(A) &2.73 \\
\object{HD\,75289} &6135 &4.43 & 0.27 &1.39$\pm$0.14 & UVES(A) &2.84  \\
\object{HD\,75289} &6135 &4.43 & 0.27 & 1.41$\pm$0.16 &  UVES(B) &2.84  \\
\object{HD\,75289} (avg) &   &   &   &1.40 &   &   \\
\object{HD\,82943} & 6025  & 4.54  &0.33  & 1.32$\pm$0.14  & UVES(A)  & 2.52   \\
\object{HD\,82943} & 6025  &  4.54 &  0.33  &  1.22$\pm$0.14  & UVES(B) & 2.52   \\
\object{HD\,82943} (avg) &   &   &   &1.27 &   &   \\
\object{HD\,92788} &5775 &4.45 &0.31 &1.13$\pm$0.16 &UVES(B) &\multicolumn{1}{c}{--} \\
\object{HD\,95128} &5800 &4.25 &0.01 &1.13$\pm$0.14 &IACUB(A) &1.71 \\
\object{HD\,108147} &6265 &4.59 &0.20 &1.02$\pm$0.14 &UVES(B) &2.34 \\
\object{HD\,114762} &5950 &4.45 &$-$0.60 &0.97$\pm$0.14 &IACUB(A) &2.26 \\
\object{HD\,117176} &5500 &3.90 &$-$0.03 &0.82$\pm$0.14 &IACUB(A) &1.76 \\
\object{HD\,120136$\dag\dag\dag$} &6420 &4.18 &0.32 &0.14$\pm$0.30 &UES &$<$1.07 \\
\object{HD\,121504} &6090 &4.73 &0.17 &1.40$\pm$0.16 &UVES(B) &2.66 \\
\object{HD\,130322} &5410 &4.47 &0.05 &1.07$\pm$0.20 &IACUB(A) &$<$0.57 \\
\object{HD\,134987} &5715 &4.33 &0.32 &1.12$\pm$0.17 &UVES(B) &$<$0.69 \\
\object{HD\,143761} &5750 &4.10 &$-$0.29 &0.95$\pm$0.19 &UES &1.30 \\
\object{HD\,145675} &5300 &4.27 &0.50 &$<$0.5 &IACUB(A) &$<$0.7 \\
\object{HD\,168443} &5555 &4.10 &0.10 &0.97$\pm$0.15 &IACUB(A) &$<$0.71 \\
\object{HD\,187123} &5830 &4.40 &0.16 &0.98$\pm$0.14 &IACUB(A) &1.20 \\
\object{HD\,195019} &5830 &4.34 &0.09 &1.27$\pm$0.15 &IACUB(A) &$<$1.05 \\
\object{HD\,210277} &5575 &4.44 &0.23 &1.05$\pm$0.17 &UVES(A) &$<$0.73 \\
\object{HD\,217014} &5795 &4.41 &0.21 &0.98$\pm$0.14 &UES &1.30 \\
\object{HD\,217107} &5660 &4.42 &0.39 &1.01$\pm$0.17 &UVES(A) &$<$0.86 \\
\hline\\[-0.15cm]
\multicolumn{7}{l}{Stars from Garc\'{\i}a L\'opez \& Perez de Taoro (\cite{Gar98}):}\\[0.1cm]
\object{HD\,75732\,A} &5150 &4.15 &0.29 &$<$0.55       &\multicolumn{1}{c}{--} & $<$0.04 \\
\object{HD\,186408  } &5750 &4.20 &0.11 &1.10$\pm$0.17 &\multicolumn{1}{c}{--} & 1.24 \\
\object{HD\,186427  } &5700 &4.35 &0.06 &1.30$\pm$0.17 &\multicolumn{1}{c}{--} & $<$0.46 \\
\hline\\[-0.15cm]
\multicolumn{7}{l}{Stars without known planets:}\\[0.1cm]
\object{HD\,870} &5425 &4.59 &$-$0.03 &0.84$\pm$0.16 &UVES(A) &$<$0.35 \\
\object{HD\,1461} &5785 &4.47 &0.18 &1.20$\pm$0.17 &UVES(A) &$<$0.71 \\
\object{HD\,1581} &5940 &4.44 &$-$0.15 &1.19$\pm$0.14 &UVES(A) &2.35 \\
\object{HD\,3823} &5950 &4.12 &$-$0.27 &0.98$\pm$0.14 &UVES(A) &2.44 \\
\object{HD\,26965A} &5185 &4.73 &$-$0.26 &0.56$\pm$0.20 &IACUB(B) &$<$0.22 \\
\object{HD\,222335} &5310 &4.64 &$-$0.10 &0.87$\pm$0.18 &UVES(A) &$<$0.35 \\
\hline\\
\end{tabular}
\newline
{$\dag$ The values for the Li abundances were taken from Israelian et al. (\cite{Isr01b}); 
we refer to this work for references. These authors used the very same set of stellar parameters
as in this paper, and thus the listed Li and Be abundances are based on the same parameter scale.
\newline
$\dag\dag$ We refer to obswww.unige.ch/$\sim$naef/who\_discovered\_that\_planet.html for the
planet discovery references.
\newline
$\dag\dag\dag$ For this star (\object{$\tau$\,Boo}) the uncertainty in $\log{N(Be)}$ is higher due to 
the ``large'' $v\,\sin{i}\sim$15\,m\,s$^{-1}$.}
\label{tab2}
\end{table*}

\subsection{Spectral synthesis}

The abundance analysis was done in standard Local 
Thermodynamic Equilibrium (LTE) using a revised version of the 
code MOOG (Sneden \cite{Sne73}), and 
a grid of Kurucz et al. (\cite{Kur93}) ATLAS9 atmospheres.
Be abundances were derived by fitting synthetic spectra to the 
data, using the same line-list as in Garc\'{\i}a L\'opez \& Perez de 
Taoro (\cite{Gar98}). While both \ion{Be}{ii} lines at 3130.420 and 
3131.065\,\AA\ are present in our data, we only used the latter, 
given the severe line blending in the region around 3130.420\,\AA\ (used 
only for checking the consistency of the fit).

In the analysis, the overall metallicity was scaled to the iron abundance. 
We then iterated by changing the Be abundance, the continuum placement and the Gaussian 
smoothing profile until the best fit for the whole spectral region was obtained 
(we fitted all the spectrum between 3129.5 and 3132.0 \AA). When considered 
important (e.g. for $\tau$\,Boo), the smoothing function used was a 
combination of a Gaussian and a rotational profile; for these cases we used 
the $v\,\sin{i}$ value determined from the width of the CORAVEL 
cross-correlation dip (Benz \& Mayor \cite{Ben84}). Three examples are shown in 
Figs.~\ref{fig1} and \ref{fig2}. The resulting abundances for all the objects 
observed are listed in Table~\ref{tab2}. Here we use the notation 
$\log{N(Be)}$=[Be]=$\log{\mathrm{(Be/H})}$+12.

\subsection{Errors}

It is not simple to derive accurate uncertainties for measurements of Be abundances 
(Garc\'{\i}a L\'opez et al. \cite{Gar95}).
In this paper the errors were estimated as follows. We considered that from 
the errors of $\pm$50\,K in temperature and $\pm$0.15\,dex in $\log{g}$ we can 
expect typical uncertainties around 0.03 and 0.06\,dex, respectively. 
{ There are several OH lines blended with the \ion{Be}{ii} 3130.420 \AA\ line; changes 
in the oxygen abundance would also change the location of the pseudo-continuum in that 
region, affecting the overall fit. To take this into account, an error 
of 0.05 dex, associated with the uncertainties in the oxygen abundances 
expected for these stars, has been added.} Other atmospheric parameters, 
like the metallicity [Fe/H] and the microturbulence, do not influence significantly 
the results, and we will conservatively consider that together they introduce an error of 0.05. 
Adding quadratically, these figures produce an uncertainty of 0.09\,dex, that was added
to the error due to continuum placement and fit quality for each case, that we conservatively
considered to be at least of 0.10\,dex. The final errors, listed in Table\,\ref{tab2} together
with the derived Be abundances, are of the order of 0.16\,dex, and quite independent of the S/N of
the spectrum. 

Note that we are interested in carrying out a differential 
analysis, and thus the knowledge
of the absolute temperatures and surface gravities is not very important. Rather,
it is crucial that these values are all in the same ``scale''.

\section{Be in stars with planets}
\label{sec:be}

There are many evidences indicating that the depletion of Be 
is connected with the rotational history of a star (e.g. Stephens et al. \cite{Ste97}). 
Although still not completely established, this history
may be related to the presence or not of a (massive) proto-planetary disk 
(Edwards et al. \cite{Edw93}; Strom \cite{Str94}; Stassun et al. \cite{Sta99}; 
Barnes et al. \cite{Bar01}; Rebull \cite{Reb01}; Hartmann \cite{Har02}). If true, this may result in 
different depletion rates for stars which had different disk masses, and thus may or not have been 
able to form the now discovered giant planets. This could, in fact, have
been the case for the pair of very similar dwarfs 
\object{16\,Cyg\,A} and B (the latter having a detected planetary companion),
for which the Li abundances seem to be quite different (Cochran et al. \cite{Coc97}; 
King et al. \cite{Kin97}; Gonzalez et al. \cite{Gon98}), but show similar amounts of Be 
(Garc\'{\i}a L\'opez \& Perez de Taoro \cite{Gar98}).
On the other hand, if pollution plays any role, we might also expect 
to detect some differences between stars with planets and stars without 
planets concerning light element abundances, and in particular
concerning Be. 

Stars with planets might thus present statistically  
different Be contents when compared with stars without (detected) planetary 
systems.

\begin{figure}[t]
\psfig{width=\hsize,file=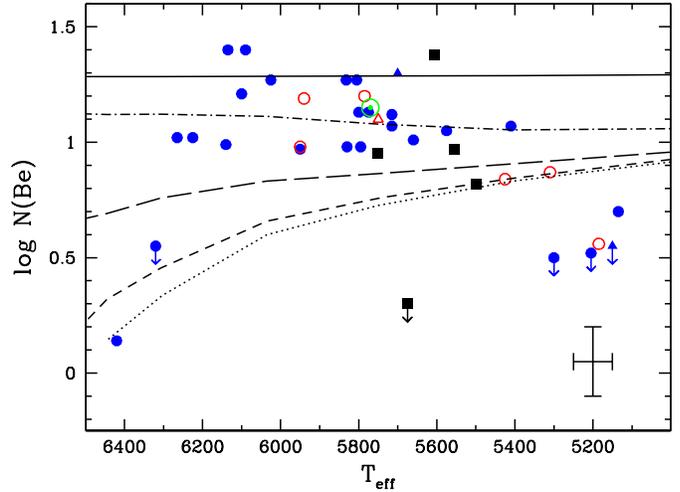}
\caption[]{Be abundances for stars with planets (filled symbols) and
stars without detected planetary companions (open symbols) as a function
of T$_\mathrm{eff}$. Circles represent the stars studied in this work,
while triangles denote the three objects taken from the study of 
Garc\'{\i}a L\'opez \& Perez de Taoro (\cite{Gar98}). Stars with
surface gravities $\log{g}\le$4.1 are denoted by squares. Superposed are the Be depletion 
isochrones (case A) of Pinsonneault et al. (\cite{Pin90}) for solar metallicity and an age of 
1.7\,Gyr. The Sun is represented by the Solar Symbol. From top to 
bottom, the lines represent a standard model (solid line), and 4 models 
with different initial angular momentum (tables 3 to 6). Typical error bars are
represented in the lower-right corner of the figure.}
\label{fig3}
\end{figure}

In Fig.\,\ref{fig3} we show a plot of Be abundances for the stars
presented in this paper as a function of T$_\mathrm{eff}$. 
Filled symbols represent stars with planets, while open symbols denote
dwarfs with no planet companions found by radial-velocity surveys (i.e. measured in
the context of the CORALIE survey (Udry et al. \cite{Udr00}), and having no clear 
radial-velocity signature of a planet).
We also included three objects from the study of Garc\'{\i}a L\'opez \& 
Perez de Taoro (\cite{Gar98}) -- triangles, namely \object{16\,Cyg\,A} (HD\,186408), 
\object{16\,Cyg\,B} (HD\,186427) and \object{55\,Cnc} (HD\,75732\,A), for which the atmospheric 
parameters have been computed using the same technique as for the stars 
presented here (Gonzalez \cite{Gon98}). In order to keep this work as an
homogeneous comparative study, we have preferred not to introduce
measurements for stars ``without'' planets taken from other works (e.g. 
Stephens et al. \cite{Ste97}), derived using different sets of atmospheric 
parameters and chemical analysis (e.g. equivalent widths).

In the figure, stars with a surface gravity $\log{g}\le$4.1 (probably 
slightly evolved) are represented by the squares. These include \object{HD\,10697}, 
\object{HD\,38529}, \object{HD\,117176}, \object{HD\,143761}, and \object{HD\,168443} 
(all planet hosts).
Note also that \object{HD\,114762}, a quite metal-poor dwarf ([Fe/H]=$-$0.6)
with a brown-dwarf companion was also included amongst the planet hosts.

Superposed with the measurements are a set of Yale beryllium depletion isochrones from 
Pinsonneault et al. (\cite{Pin90}) for solar metallicity and an age of 
1.7\,Gyr. Given that it is not possible to know what was the initial 
Be abundance for each star, we have considered an initial 
$\log{N(Be)}=1.28$, i.e., between Solar ($\log{N(Be)}$=1.15 -- Chmielewski et al. \cite{Chm75}) 
and meteoritic ($\log{N(Be)}$=1.42 -- Anders \& Grevesse \cite{And89}).

Although still preliminary, a look at the figure shows that the current results argue
against pollution as the key process leading to the metallicity excess of stars with planets
(see also Santos et al. \cite{San01}, \cite{San01b} and Pinsonneault et al. \cite{Pin01}). 
Adding $\sim$50 earth masses of C1 chondrites
to the Sun, for example, would increase its iron abundance by about 0.25\,dex 
(a value similar to the average difference observed between stars with and without detected giant 
planets), and its Be abundance would increase by a slightly higher factor\footnote{For this 
calculation we have used a Be/Fe ratio similar to C1 chondrites, and considered an iron content of 
$\sim$20\% by mass (Anders \& Grevesse \cite{And89}). Note also that the fall of a few 
jupiter-mass jovian planet(s) would not have a big effect on the iron or beryllium 
abundance of the star.}. No such difference seems present in our data.
In the same way, the results also do not support extra-mixing due to
an eventual different angular momentum history of the two ``populations'' of stars. 
We note, however, that at this stage we are limited by the small number of
comparison stars analyzed in this work.

In this context, one particular star (\object{HD\,82943}) deserves a few comments. This planet host
was recently found to have a near meteoritic $^6$Li/$^7$Li ratio (Israelian et al. \cite{Isr01a}), 
better interpreted as a sign that planet/planetary material was engulfed
by this dwarf sometime during its lifetime. Israelian et al. have
suggested that the Li isotopic ratio of \object{HD\,82943} could be explained by the fall of 
either the equivalent of a 2 jupiter mass giant planet or a 3 earth mass terrestrial planet (or 
planetary material). Considering that \object{HD\,82943} had initially a solar Be abundance, 
the addition of this quantity of material would increase its Be abundance by about 0.1\,dex. A small
``excess'' of Be would thus be observed, but this value would be within the error bars of our 
measurements. As we can see from Fig.\,\ref{fig3},
the Be abundance of \object{HD\,82943} is not particularly high when compared to other stars in the plot
(although it seems to occupy a position near the upper envelope of the Be abundances). No further 
conclusions can be taken.

\subsection{The $\log{N(Be)}$ vs. T$_\mathrm{eff}$ trend}

According to the standard models (without rotation -- e.g. Table\,1 of Pinsonneault 
et al. \cite{Pin90}), basically no Be depletion should occur for dwarfs in the temperature 
region plotted in the figure. This is clearly not compatible with the observations. 
However, it is already known that standard models are not able to explain most of the observed behaviors
of Li and Be (e.g. Stephens et al. \cite{Ste97}). On the other hand, models with rotation 
(Pinsonneault et al. \cite{Pin90})\footnote{Improved Yale models 
(e.g. Deliyannis \& Pinsonneault \cite{Del93}) do not show an 
important difference. The 1990 models were thus used in the current paper, 
since they are the only tabulated.} were 
shown to be quite satisfactory for stars with 6500\,K\,$\le$\,T$_\mathrm{eff}$\,$\le$5600\,K 
(Stephens et al. \cite{Ste97}). But as we can see from the figure, they do not 
predict significant burning at temperatures around 5200\,K. We note that according to these 
isochrones, Be depletion is done at moderate rates once the star reaches the main-sequence (about 0.06\,dex 
from 0.7 to 1.7\,Gyr for a 0.9\,M$_{\sun}$ dwarf, and at a considerably lower rate for higher 
ages following the Fig.\,11 of Stephens et al. \cite{Ste97}), and thus we do not expect a
crucial difference with the presented curves. 

In contrast with the models\footnote{We will assume for now that 
the initial Be abundance was similar (within the uncertainties) 
for all the stars in the plot.}, one very interesting detail is seen.
If, for temperatures between about 5600 and 6200\,K, the dispersion
in the $\log{N(Be)}$ is remarkably small, and the abundances seem to be
consistent with the model predictions, the general trend 
for lower temperatures follows a 
slow decrease with decreasing temperature. Under the assumption that no
difference exists between stars with and without giant planets, either the trend is due to some 
metallicity or age effect, or it is simply telling us that the models are not able to reproduce 
the observations for temperatures bellow $\sim$5600\,K. This
problem was also noted by Stephens et al. (\cite{Ste97}), but no such
extreme cases had been found, maybe because their ``solar metallicity sample'' 
did not go down to temperatures lower than $\sim$5500\,K. 
Note that the iron abundances for the stars in the plot are in the range 
from $-$0.6 to $+$0.5, and the variation of the initial 
Be abundance with stellar metallicity seems to be quite small in this interval
(Rebolo et al. \cite{Reb88}; Molaro et al. \cite{Mol97}; Garc\'{\i}a L\'opez 
et al., in preparation). 

Particularly interesting are the 4 cases for which only upper values for the Be 
abundances were found: 
\object{HD\,38529}, \object{HD\,145675} (\object{14\,Her}), \object{HD13445} 
(\object{Gl\,86}), and \object{HD\,75732} (55\,Cnc), all in the temperature 
interval between 5150\,K and 5700\,K. A detail of the spectral synthesis for one of these
stars can be seen in Fig.\ref{fig2}.
While for \object{HD\,38529} the Be depletion may be explained by the fact 
that this star is already leaving the main-sequence (its low 
surface gravity and its position in the HR-diagramme -- e.g. Gonzalez et al. \cite{Gon01} -- show 
it to be evolved) for the three remaining objects no simple explanations seem to exist. 

One possible explanation for Be abundances of the
most discrepant objects could be their ages. For example, \object{55\,Cnc} and
\object{Gl\,86} seem to be quite old, with isochrone ages in agreement with the oldest stars in the
galaxy (ages were determined from theoretical isochrones of  Schaller et al. \cite{Sch92}, 
and Schaerer et al. \cite{Sch93a}, \cite{Sch93b}). This is not 
the case for \object{14\,Her}, with an age around $\sim$6\,Gyr. 
Furthermore, for \object{14\,Her}, there is a star (\object{HD\,222335}, 
without detected giant planet) with about the same temperature, 
similar age ($\sim$5-6\,Gyr) but much higher Be abundance. 

\begin{figure}[t]
\psfig{width=\hsize,file=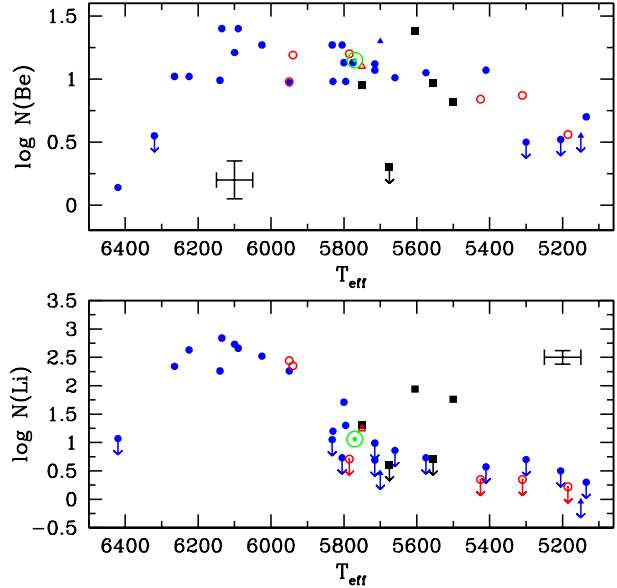}
\caption[]{Lithium and Beryllium abundances for the stars studied in this paper. 
We note that for some of the stars, no Li abundances were found in the literature 
(see Table\,\ref{tab2}). Symbols as in Fig.\,\ref{fig3}. Typical error bars are
represented in each panel (0.16 for Be and 0.12 for Li).}
\label{fig4}
\end{figure}

Can a metallicity effect explain the observed difference? An increase in the metallicity
(and thus in the opacities) has the effect of changing the convective envelope depth of a
star. It has been shown by Swenson et al. (\cite{Swe94}) that
even small changes in the oxygen abundances may have a strong effect on the Li depletion rates.
Although it is unwise to extrapolate directly, a similar effect should exist concerning Be.
In what concerns our case, most of the objects in the plot having temperatures lower
than 5400\,K are metal-rich with respect to the Sun. For example, \object{14\,Her} is one of the 
most metal rich stars in the sample ([Fe/H]=$+$0.50). But some counter examples
exist, like \object{Gl\,86}, only about 100\,K cooler, with a value of [Fe/H]=$-$0.20.
Unfortunately, no theoretical isochrones for Be depletion are available 
for these high metallicities, and it is not clear whether a difference of 
$+$0.5\,dex in metallicity can or cannot induce significantly different 
Be depletion rates for this temperature range. { On the other hand,
if some unknown opacity effect could lead to systematic errors as a function of
temperature -- e.g. the missing UV opacity (Balachandran \& Bell \cite{Bal98}) --
we could maybe expect to measure some trend. However, we have no special reason to believe 
that this effect can lead to the observed trend. Furthermore, the missing UV opacity problem is
far from being solved (if it exists at all); indeed, recent results seem to show that we can 
obtain a good fit of the solar spectrum without taking into account this extra-opacity (Allende 
Prieto \& Lambert \cite{All00}).  }

Although not conclusive, the evidences discussed above 
suggest that a {\it bona-fide} explanation for the observed Be abundances 
may pass either by some inconsistency in the models, or by some 
metallicity effect. But given the much higher number of planet hosts in the plot when
compared with the non-planet hosts, we still cannot discard that the presence
of a planet might be the responsible for the observed trend\footnote{Can the 
presence of a massive proto-planetary disk (able to
form planets), or other variables connected to the process of planetary 
formation itself (e.g. timescale), induce different angular momentum histories 
for stars with different T${_\mathrm{eff}}$?}.
Else, since Be depletion in the Pinsonneault 
et al. (\cite{Pin90}) models is strongly related to the angular momentum lost,
we would have, for example, to consider that there is some negative correlation between 
initial angular momentum and/or angular momentum loss and stellar mass (which does 
not seem very plausible).
Unfortunately only the addition of more objects to the plot, and in particular 
the determination of Be abundances for a larger sample of dwarfs with 
T$_\mathrm{eff}<$5400\,K will permit to better settle down this question.
Such observations are currently in progress.

It is also worth mentioning that on the other side of the T$_\mathrm{eff}$ 
plane, both \object{HD\,120136} (\object{$\tau$\,Boo}) and \object{BD\,$-$10\,3166}, positioned 
in the Li (and Be) dip region (e.g. Boesgaard \& King \cite{Boe01b}), have a Be abundance 
that may be compatible with the models.

\subsection{Lithium vs. Beryllium}

{ 
Our data can also be used to further investigate the issue of Li and Be depletion in
main sequence stars. Fig.\,\ref{fig4} compares Li and Be abundances for the same stars, and
shows, as expected, that Li is burned much faster than Be, and its decline 
with temperature is much more clear. This is clearly 
expected from the models for a middle age solar type dwarf in this 
temperature regime. Only two points seem to come out of the main trend { (in the Li panel)}. 
These are the cases of \object{HD\,10697} and \object{HD\,117176} (70\,Vir). These two stars
deserve some particular discussion in Sect.\,\ref{sub2stars}.

In Fig.\,\ref{fig5} we plot Be vs. Li abundances for all the stars
with T$_{\mathrm{eff}}<$6300\,K for which both Li and Be abundances are available.
The plot reveals no clear trends. Even \object{70\,Vir} and \object{HD\,10697} follow the 
main trend, probably attesting the normality of these two stars. 
Again, no significant difference is seen 
between the two groups of stars (planet hosts and non-planet hosts).

One interesting but expected detail in Fig.\,\ref{fig5} is that there are no stars 
in the lower right part of the plot, a region that would correspond to stars having depleted 
much of their Be but that are still Li rich. 
This ``gap'' in the figure, plus the few stars in the upper right corner, suggests that there 
might be a correlation between Li and Be depletions (like the one found by Deliyannis 
et al. \cite{Del98}, Boesgaard et al. \cite{Boe01} and more recently 
by Boesgaard \& King \cite{Boe01b} for their sample of hotter dwarfs, but absent in the
study of Dom{\'\i}nguez Herrera \cite{Her98}, and in the sample of old cluster members 
of Randich et al. \cite{Ran01}). However, there is a significant number of stars in the upper left 
region of the plot having Be determinations but only upper limits for the Li abundance (objects 
that are not present in Fig.\,14 of Boesgaard et al.). 
This particular point in quite interesting,
since it is, to our knowledge, the first time such a ``population'' of stars is discussed
in the literature.

\begin{figure}[t]
\psfig{width=\hsize,file=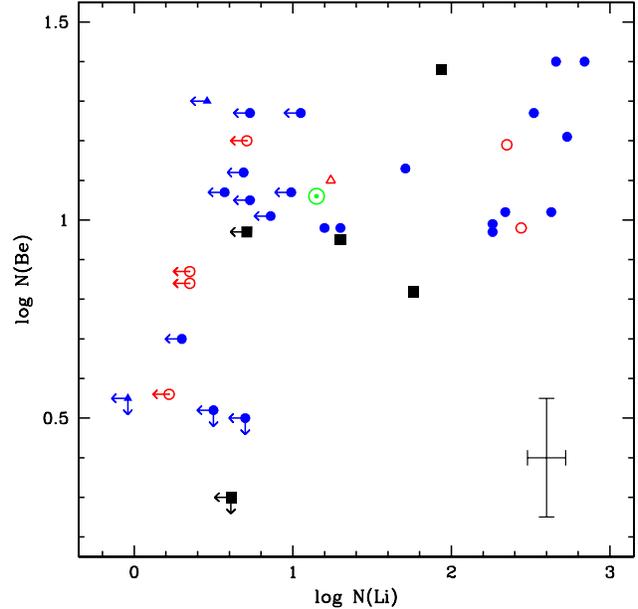}
\caption[]{Li vs. Be abundances for the stars in our sample having temperatures lower than 6300\,K and for which both Li and Be abundances were available. Symbols as in Fig.\ref{fig3}. Typical error bars are represented in the lower-right corner of the figure.}
\label{fig5}
\end{figure}

\begin{figure}[t]
\psfig{width=\hsize,file=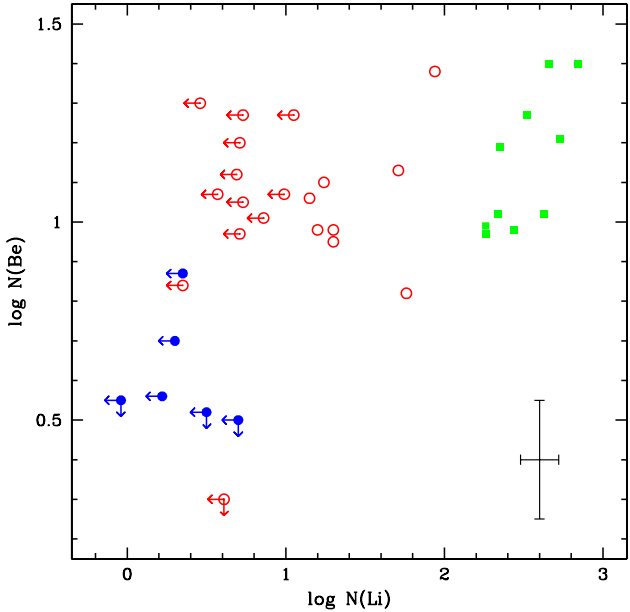}
\caption[]{Same as Fig.\,\ref{fig5} but using different symbols for three temperature intervals:
T$_{\mathrm{eff}}$$\le$5400\,K (filled circles), 5400\,K$<$T$_\mathrm{eff}$$\le$5900\,K (open circles), and
T$_{\mathrm{eff}}$$>$5900\,K (filled squares). See text for more details.}
\label{fig6}
\end{figure}

In Fig.\,\ref{fig6} we have done the same plot as in Fig.\,\ref{fig5}, but this time
using different symbols as a function of the temperature of the object. 
As we can easily see from the plot, the objects in the three
different temperature regimes chosen occupy clearly different positions in the diagramme. Stars
with T$_{\mathrm{eff}}$$\le$5400 are positioned in the lower-left corner. These correspond basically
to objects having burned some Be, and that only have upper values for the Li abundance.
In the upper-right corner of the figure are positioned stars in the range 
T$_{\mathrm{eff}}$$>$5900\,K. These correspond to objects that have essentially not burned any
Be, having depleted only moderately their Li content.
An intermediate population, with 5400\,K$<$T$_{\mathrm{eff}}$$\le$5900\,K, having
already considerably depleted its original Li but not its Be content is present in the mid-upper 
part of the plot.

Except for \object{HD\,38529},
a sub-giant star (the object with the lowest Be abundance in the plot), all the other objects are well grouped according to their temperature range.
This is an interesting result, since it is giving us important clues about the temperature at which 
the onset of Li and Be depletion occurs in our metal-rich stars: if for Li depletion occurs already 
for temperatures around 5900\,K, for Be this only seems to happen at T$_{\mathrm{eff}}$$\sim$5400\,K.
Note, however, that given the dispertion these limits have to be seen as approximate.

Note also that we did not include \object{$\tau$\,Boo} 
in Figs.\,\ref{fig5} and \ref{fig6}, since its the only object in the sample that is 
clearly in the Li-dip region. 

}

\subsubsection{\object{HD\,10697} and \object{70\,Vir}}
\label{sub2stars}

{ As mentioned above, \object{HD\,10697} and \object{70\,Vir} do not fit into
the mean trend in the lower panel of Fig.\,\ref{fig4}.
Considering as initial abundances of Li the ``cosmic'' value
($\log{N(Li)}$=3.31; Martin et al. \cite{Mar94}) and for Be the average between
meteoritic and solar ($\log{N(Be)}$=1.28, as above)}, then the observed 
values for \object{70\,Vir} indicate that Be ($\log{N(Be)}$=0.82)
is depleted { about 3 times}, while Li ($\log{N(Li)}$=1.76) is depleted by 
by a factor of 22. Nevertheless, such a pattern of Li and Be depletion
is not so unexpected (see e.g. Deliyannis \& Pinsonneault \cite{Del97}; 
Stephens et al. \cite{Ste97}), and is compatible with models of 
depletion including turbulent induced rotational mixing (see e.g. discussion 
in Stephens et al. \cite{Ste97}). But \object{70\,Vir} is not a young object 
(it seems to be slightly evolved, as confirmed by its low 
surface gravity $\log{g}$=3.90), and it seems strange that an evolved 
star with its temperature may still have so much Li.

One possibility would be to invoke Li and Be dredge up from a ``buffer''
below the former main-sequence convective envelope (Deliyannis et al. \cite{Del90}).
But as discussed in e.g. Randich et al. (\cite{Ran99}), this scenario does not
seem to be supported by the observations.
However, the fact that this star is evolved means that its 
current temperature is probably different to the one it had
during the main-sequence phase. With a near solar metallicity ([Fe/H]=$-$0.03) and a 
mass of $\sim$1.1\,M$_{\sun}$, the temperature of 
\object{70\,Vir} would have been more close to 6000\,K. A dwarf with this temperature is 
not supposed to burn much of its Li (as we can see from the plot).
So, we speculate that this star may have just left the main-sequence,
and started to dilute (and/or burn) the Li (as well as Be) content present in its
convective envelope. If the size of the convective envelope 
has increased ``quite fast'', maybe it still did not have time to deplete/dilute all 
the Li, but is already depleting part of its Be (Charbonnel et al. \cite{Cha00}).

The case of \object{HD\,10697} is in fact not very different from the one of
\object{70\,Vir}. Both have the same mass (1.10\,M$_{\sun}$), and similar surface gravities. 
The main difference resides in the fact that \object{HD\,10647} seems to have, { within the 
errors}, preserved most of its Be content, while Li is 
depleted about 23 times. But given the uncertainties in our Be abundance determination,
and the unknown initial Li and Be content of the star, { we cannot be certain 
whether} the Be is really intact in the atmosphere of this object. The same arguments 
discussed for 70\,Vir { may thus be} valid in this case.

{ It is worth mentioning that another possibility to explain the high Li abundances of 
these two stars would be to invoke a 
planet engulfment} (like in the case of \object{HD\,82943} 
-- Israelian et al. \cite{Isr01a}), that could simply result from the
migration of a former close-in planet due to tidal interactions with the evolved star 
(Rasio et al. \cite{Ras96}). Unfortunately, no measurements of the Li isotopic ratio are available 
for \object{70\,Vir}. But recent models by Siess \& Livio (\cite{Sie99}) show that the 
absorption of a planet by a giant star would have the effect of increasing 
considerably its Li abundance, but only slightly the Be content, a situation that could be 
compatible with the observed Li and Be abundances for this star.

{ There is another interesting point about these two stars. Although they have
very similar Li abundances, their Be contents differ by 0.56\,dex. Given that
both stars have similar metallicity ($-$0.03 and $+$0.16, for 70\,Vir and HD\,10697, 
respectively) their initial Be content is not expected to be significantly different 
(Rebolo et al. \cite{Reb88}; Molaro et al. \cite{Mol97}; Garc\'{\i}a L\'opez 
et al., in preparation). However, we do not know up to which extent Be and Li abundances are
uniform in the Inter-Stellar Medium. A different initial Li/Be ratio, 
together with the uncertainties in the Li and Be determinations,
could probably lead to the observed effect. Else, we would need to invoke that
the two stars have burned Li at the same rate, while the opposite happened for Be. 
But given that these two stars are sub-giants, we cannot
exclude that we are simply observing some evolutionary effect, eventually combined
with slightly different initial Li and Be abundances. Note that
the former object is a bit cooler, being positioned in the temperature 
region where the Be depletion seems to occur (see previous Sect.).}

\section{Conclusions}
\label{sec:conclusions}

We have obtained Be abundances for a sample of planet hosts stars, and a smaller
sample of stars without known giant planetary companions, aimed at studying the existence 
of any significant difference between the two groups. { Our data also allow us to further 
investigate Li and Be depletion among metal-rich stars.} The objects have metallicities
between about $-$0.6 and $+$0.5\,dex, and cover the temperature interval between
$\sim$5150\,K and 6450\,K. The main conclusions go as follows:

\begin{itemize}

\item No particular difference was found between the Be abundances of stars with and without 
giant planetary companions in the temperature interval for which both groups were observed.
This may be interpreted as an argument against the ``pollution scenario'' as the
source of the high-[Fe/H] observed for planet host stars. However, given the low
number of comparison stars available in this study, no strong constraints can still
be set.

\item For stars with T$_\mathrm{eff}<$5500\,K, the general trend of the Be 
abundances is not well described by models of light element 
depletion for solar metallicity (e.g. Pinsonneault et al. \cite{Pin90}). 
These inconsistencies may be connected either to some metallicity effect, or
to some problem with the models, but at the moment we cannot completely discard some 
relation with the presence of a planet.

\item For temperatures lower than 6300\,K, no clear correlated 
Li and Be pattern is seen, contrary to what was found by other 
authors for hotter dwarfs (Deliyannis et al. \cite{Del98}; 
Boesgaard et al. \cite{Boe01}).

{
\item Our data permit to establish approximate temperature limits for 
the onset of strong Li and Be depletion for the metal-rich stars studied in
this work.
For stars with T$_\mathrm{eff}<$5900\,K, severe Li burning is seen, while for
Be this limit is around 5400\,K.
}

\end{itemize}

A large survey for Be in planet and non-planet host stars is currently under way, 
and we expect it to give new insights into this timely and interesting subject.

\begin{acknowledgements}
  We thank our anonymous referee for the useful and
  interesting comments and suggestions.
  We wish to thank the Swiss National Science Foundation (FNSRS) for
  the continuous support to this project. 
  This research was also partially supported by the Spanish DGES under
  project PB98-0531-C02-02.  Support from Funda\c{c}\~ao para a Ci\^encia e Tecnologia, 
  Portugal, to N.C.S. in the form of a scholarship is gratefully acknowledged.
\end{acknowledgements}

%---------------------------bibliography---------------------------

\end{document}